\documentclass{pasj02}

\jyear{2025}
\Received{2024/10/28}
\Accepted{2025/06/08}
\Published{2025/08/XX}

\begin{document} 

\title{ The July 2023 Outburst of Comet 12P/Pons-Brooks: Observations and
Modeling of Dust Coma and Arc Structure }

\author{
 Hitoshi \textsc{Hasegawa},\altaffilmark{1}\altemailmark\orcid{0009-0009-6913-4780} \email{kin3.hasegawa@gmail.com} 
 Mitsunori \textsc{Tsumura},\altaffilmark{2}\orcid{0000-0000-0000-0000}
 Shin-ichi \textsc{Watanabe},\altaffilmark{3}
 Hiroki \textsc{Akisawa},\altaffilmark{4}
 and 
 Jun-ichi \textsc{Watanabe}\altaffilmark{5}\orcid{0000-0000-0000-0000}
}
\altaffiltext{1}{2-19-18-304 Kamitakaido Suginami-ku, Tokyo 168-0074, JAPAN}
\altaffiltext{2}{Wakayama-shi, Wakayama 640-0351, JAPAN}
\altaffiltext{3}{Niigata-shi Niigata 950-2162, JAPAN }
\altaffiltext{4}{Taishi-cyo, Hyogo 671-1511, JAPAN}
\altaffiltext{5}{National Astronomical Observatory of Japan, 2-21-1 Osawa Mitaka Tokyo 181-8588, JAPAN}



\KeyWords{comets: outburst --- comets:dust --- comets:rotation axis}  

\maketitle

\begin{abstract}

We conducted continuous imaging observations of Comet 12P/Pons-Brooks, which experienced multiple outbursts since July 2023. Among these, we determined the dust coma expansion velocities and the occurrence times for three outbursts. The outburst on July 20.39$\pm$0.05, 2023 had an expansion velocity of 207.5$\pm$0.1 $\rm{m\,s}^{-1}$. The outburst on November 14.52$\pm$0.10, 2023 had an expansion velocity of 305.8$\pm$13.7 $\rm{m\,s}^{-1}$, and the outburst on April 2.23$\pm$0.06, 2024 had an expansion velocity of 408.9 $\pm$14.3$\rm{m\,s}^{-1}$.
During the July outburst, in addition to the isotropically expanding dust coma, a semicircular structure (arc structure) was observed. This structure expanded without changing shape over two months, until mid-September, at a slower velocity of 52.0$\pm$1.7$\rm{ms}^{-1}$ compared to the outer coma. We hypothesize that this arc structure consists of dust ejected over a certain period during the outburst from an active region at mid-latitudes on the comet's nucleus, with a rotation axis near the line of sight. We performed model calculations to reproduce the structure.
As a result, we found that the observations were well reproduced by assuming a rotation axis direction of RA=$85\degree$, Dec=$-30\sim-35\degree$ and an active region latitude of $70\degree$ during the July outburst. The active period of the outburst was shorter than the rotation period, approximately 0.7 times the rotation period. Since the dust particles comprising the arc structure escaped dissipation due to solar radiation pressure, we infer that they are dust with a ratio of radiation pressure to gravity of $\beta\sim0.1$ which corresponds to a dust particle size of about 10 $\micron$.

\end{abstract}

\section{Introduction}

  Comet 12P was originally discovered by J. L. Pons on 1812 July 21 at Marseille, France. It was followed during that apparition until September 28. While the observation period was only two months, its orbit was poorly estimated to be a period of about 70 years by J. F. Encke. Actually, in July 1883, W. R. Brooks discovered a new comet which was later identified with the comet of 1812. This comet is therefore called as 12P/Pons-Brooks. It should be noted that this comet in 1883-1884 was well observed with several outbursts. One occurred in September 22-23, and the strongly bright central condensation gradually diffused and became darker. The comet became bright enough to be seen by naked eye in November 1883, when some strange T-shape of the coma was reported by telescopic observation \citet{young}. Another outburst occurred in beginning of 1884. After the perihelion passage of January 26.22, 1884, this comet was observed until June. The long duration of the observation leaded to the orbit determination with high precision. Based on this orbit, the next return of this comet was predicted, and recovered in June 20, 1953, while the perihelion passage would be May 22, 1954. After the recovery, this comet displayed several outbursts as same as previous apparition. The large outburst occurred on July 1st, when the magnitude of the comet became 13th from 18th. Another was occurred in March 1954. It is clear that this comet has tendency to have occasional outbursts inbound toward the perihelion passage. The details of all three apparitions can be found in \citet{kronk-a} and \citet{kronk-b}.

  The next perihelion passage was estimated at April 21, 2024, while it was recovered at June 20, 2020 at the large heliocentric distance of 11.9 au by Discovery Telescope at the Lowell Observatory. When it came to the heliocentric distance of 3.9 a.u., a large outburst was occurred on July 20, \citep{green}. The brightness dramatically increased, with its magnitude changing from 16-17 to 11-12 and accompanied with curious morphological change such as a strange horseshoe shape. Hereafter, this horseshoe-shaped structure will be referred to as an arc structure. Since then, there have been several outbursts toward the perihelion passage. 

Observational studies have been conducted on several comets that have
exhibited outbursts. \citet{sekanina} demonstrated that the jet
structures of comet 81P/Wild 2 consist of dust ejected from specific
regions on the nucleus. Using coma morphology observations from the
Deep Impact mission of comet 9P/Temple 1, an attempt was made to
pinpoint the locations of the jets and to estimate their influence on
the comet’s rotation \citep{farnham}.

Comet 29P/Schwassmann–Wachmann, well known for its frequent outbursts,
has been extensively studied. \citet{huges} found that the outbursts
of comet 29P reached maximum brightness on average 2.5 days after
initiation, after which they gradually diffused over a period of 20–30
days, with expansion velocities ranging between 100 and 400 m/s. These
brightness variations and expansion speeds were similar to those
observed for comet 12P/Pons–Brooks in our study.

\citet{trigo-rodriguez} proposed that the crystallization of
amorphous ice is the most likely mechanism for the outbursts of comet
29P. In the case of comet 17P/Holmes, which underwent a large-scale
outburst in 2007, \citet{montalto} attributed the event to
internal instabilities, such as the collapse of internal voids and the
release of subsurface volatile materials.

There are several hypotheses for cometary outbursts.  One theory
attributes these events to internal processes, like the energy release
from the crystallization of amorphous water ice \citep{prialnik}.
Another theory considers external factors, such as collisions
with interplanetary objects or the influence of solar radiation
\citep{huebner}. Additionally, the physical structure of the
comet's nucleus, including internal cavities and phase transitions, is
also proposed as a possible cause \citep{gronkowski}.  However, there
is currently no established theory regarding cometary outbursts. Since
there are few examples where the outburst phenomenon can be tracked
morphologically, as in the case of Comet 12P, our analysis may
contribute to the understanding of this phenomenon.

The Rosetta mission provided detailed observations of small-scale dust
jets on the nucleus of comet 67P/Churyumov–Gerasimenko. \citet{pajola}
demonstrated that landslides can trigger dust release,
identifying traces of cliff collapses—presumed sources of dust
emission—on the cometary surface. These locations exhibited exposures
of fresh, high-albedo internal ice. A physical model of the dust
ejection mechanism, based on the sublimation of subsurface volatile
ices through the development of cracks in the cliffs, was developed by
\citet{skorov}. \citet{vincent} revealed that the
positions of the dust jets correlate with steep cliffs. At present,
the relationship between these small-scale dust jets and the
large-scale outbursts remains unclear.
High-resolution imaging by the Rosetta mission enabled quantitative
observations of outbursts on comet 67P/Churyumov–Gerasimenko \citep{lin}.
The authors identified possible causes of the outbursts,
including thermal stress, seasonal heating, cliff collapses, and the
release of volatiles from subsurface fractures.  They suggested that
these phenomena are likely not caused by a single mechanism, but
rather by a combination of multiple factors.

  In this paper, our results of the numerical simulation for reconstructing the arc structure, observed from July to September 2023, as transient ejection of the dust particle of appropriate size from one discrete source on the nucleus.

\begin{figure}\label{fig:1}
 \begin{center}
   \includegraphics[width=8.5cm]{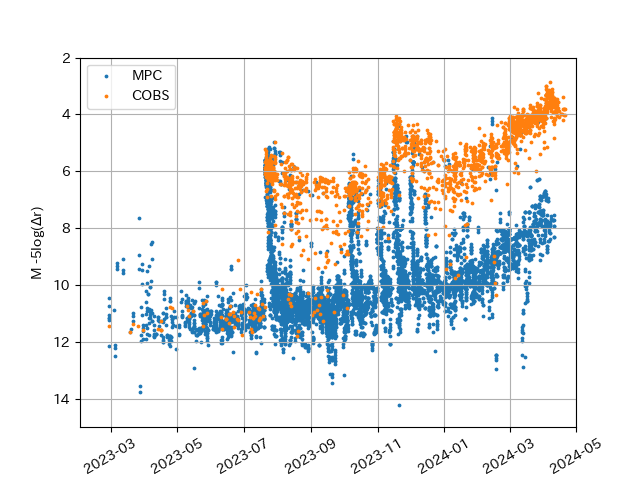} 
 \end{center}
 \caption{Light curve of Comet 12P/Pons-Brooks reported in MPC and COBS. Heliocentric and geocentric distances are corrected. {Alt text: Light curve of Comet 12P/Pons-Brooks}}
\end{figure}

\section{Observations and measurements}\label{sec:2}

The first report of an outburst during the current return of Comet 12P/Pons-Brooks was made by F. Kugel (Dauban, France), who reported it as having a magnitude of 11.5 on Jul 20.90 UT, 2023 \citep{green}. Prior to this, the brightness of the comet was around magnitude 16 to 17, and the outburst caused an abrupt increase in brightness by approximately 5 magnitudes.

\begin{table}
  \tbl{Observation devices. D is the aperture of the telescope and FL is the focal length.}{
  \begin{tabular}{crrll}
      \hline
      Observer & D & FL & Detector & Obs. site  \\ 
               & (mm) & (mm) & &  \\ 
      \hline
      M.Tsumura & 300 & 1200 & ZWO ASI6200 Pro & Wakayama Japan \\
      S.Watanabe & 356 & 3910 & ZWO ASI290MM & Niigata Japan \\
      \hline
  \end{tabular}}\label{tab:1}
\end{table}

Figure 1 shows the brightness observations of Comet 12P reported to the MPC (Minor Planet Center \footnote{<https://minorplanetcenter.net/mpcops/submissions/cometary/>}) and COBS (Comet Observation Database \footnote{<https://www.cobs.si/obs/comet/>}), plotted after corrections for geocentric and heliocentric distances. Since MPC mainly focuses on precise positional observations, the magnitudes are generally fainter due to narrow-aperture photometry. On the other hand, COBS reports are aimed at measuring the total brightness, so the reported magnitudes tend to be brighter, with a difference of around 4 magnitudes between the two sources.

However, from Figure 1, we can see that a total of seven outbursts, both large and small, were recorded between July 2023 and April 2024. The brightness increases due to the outbursts were around 4 to 5 magnitudes. Notably, the outbursts in July and November 2023 were particularly prominent.

We conducted continuous observations over a period of 10 months following the report of the outburst on July 20, 2023 tracking the expansion and structural changes of the dust coma near the nucleus. The telescopes, detectors, and observation sites used for the observations are listed in Table 1. During the observation period, we were able to obtain continuous imaging observations for the outbursts that occurred in mid-July 2023, mid-November 2023, and early April 2024, allowing us to study the structure and expansion of the dust coma during each outburst.

The observation list for the July 2023 outburst is summarized in Table 2. The table also includes measurements of the radius of the outer edge of the dust coma, as well as the radius of the semicircular arc structure (hereafter referred to as the arc structure) observed in the images. After the first outburst, several small-scale outbursts occurred, followed by a large outburst observed on November 14, 2023. The observation list for the November outburst is summarized in Table 3. After a quiet period from January 2024 onward, another outburst was observed in early April 2024. The observation list for the April 2024 outburst is summarized in Table 4.

\begin{figure*}\label{fig:2}
 \begin{center}
   \includegraphics[width=15cm]{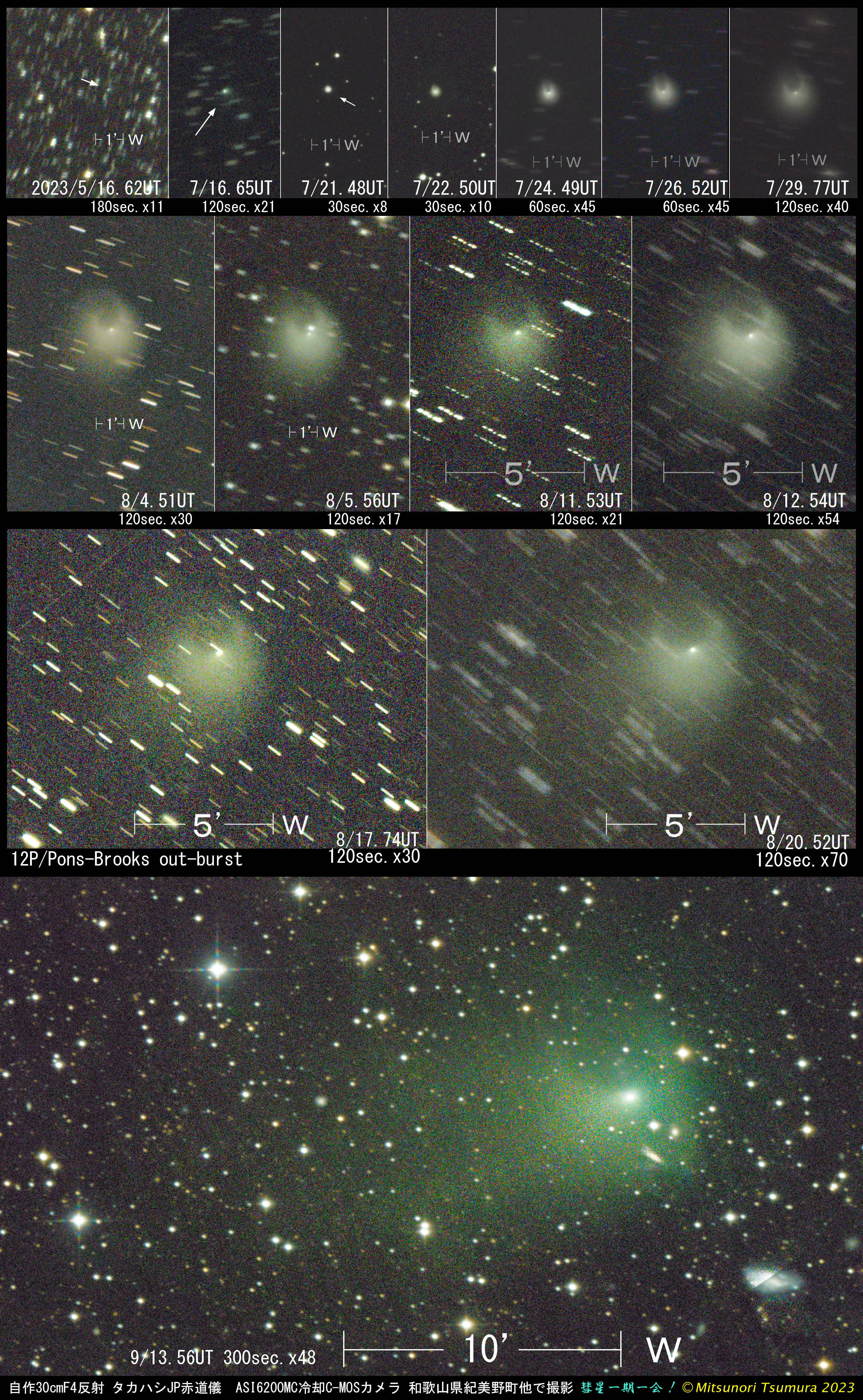} 
 \end{center}
 \caption{Development of the arc structure observed after July 20, 2023 outburst. 
The arc structure remained even on September 13, 2023, two months after the outburst.
{Alt text: Observed images of Comet 12P outburst}}
\end{figure*}

\begin{table}
  \centering
  \caption{Observational results of the outburst started on July 20, 2023. W in Obs. column denotes Watatanabe, T denotes Tsumura.}{
\begin{tabular}{crrrrl}
\hline
DATE(UT)     & r & $\Delta$ & Coma & Arc & Obs. \\
             & (au) & (au) & (km) & (km) & \\
\hline
2023-07-21.63 & 3.87 & 3.56 & 19100.0 & 5000.0 & W \\
2023-07-22.50 & 3.86 & 3.55 & 50000.0 & 13000.0 & T \\
2023-07-22.63 & 3.86 & 3.55 & 38100.0 & 11000.0 & W \\
2023-07-23.51 & 3.86 & 3.54 & 52500.0 & 15000.0 & W \\
2023-07-24.49 & 3.84 & 3.54 & 80000.0 & 23000.0 & T \\
2023-07-24.63 & 3.84 & 3.53 & 80000.0 & 22000.0 & W \\
2023-07-25.50 & 3.83 & 3.53 & 73850.0 & 29000.0 & W \\
2023-07-26.52 & 3.82 & 3.52 & 94950.0 & 37000.0 & T \\
2023-07-26.63 & 3.82 & 3.52 & 110000.0 & 35000.0 & W \\
2023-07-27.71 & 3.81 & 3.51 & 97700.0 & 39000.0 & W \\
2023-07-28.57 & 3.80 & 3.50 & 126500.0 & 47000.0 & W \\
2023-07-29.77 & 3.79 & 3.50 & 170000.0 & 49000.0 & T \\
2023-08-04.51 & 3.74 & 3.46 & 280000.0 & 67000.0 & T \\
2023-08-05.56 & 3.72 & 3.45 & 300000.0 & 69000.0 & T \\
2023-08-11.53 & 3.66 & 3.41 & 370000.0 & 103000.0 & T \\
2023-08-12.54 & 3.65 & 3.41 & 425000.0 & 102000.0 & T \\
2023-08-17.74 & 3.60 & 3.37 & 500000.0 & 123000.0 & T \\
2023-08-20.52 & 3.57 & 3.36 & 570000.0 & 158000.0 & T \\
2023-09-13.56 & 3.31 & 3.22 & - & 208000.0 & T \\
\hline
\end{tabular}}\label{tab:2}
\end{table}

\begin{table}
  \centering
  \caption{Same as Table 2, but for the outburst of Nov 14, 2023.}{
\begin{tabular}{lrrrl}
\hline
DATE(UT)     & r & $\Delta$ & radius & Obs. \\
             & (au) & (au) & (km) &  \\
\hline
2023-11-19.43 & 2.53 & 2.72 & 146500.0 & T \\
2023-11-20.41 & 2.52 & 2.71 & 165500.0 & T \\
2023-11-22.39 & 2.49 & 2.69 & 211500.0 & T \\
2023-11-23.40 & 2.48 & 2.68 & 234000.0 & T \\
2023-11-26.40 & 2.45 & 2.65 & 312000.0 & T \\
2023-11-15.47 & 2.58 & 2.77 & 19000.0 & W \\
2023-11-20.48 & 2.52 & 2.71 & 142500.0 & W \\
2023-11-21.43 & 2.51 & 2.70 & 175500.0 & W \\
2023-11-22.40 & 2.49 & 2.69 & 208500.0 & W \\
\hline
\end{tabular}}\label{tab:3}
\end{table}

\begin{table}
  \centering
  \caption{Same as Table 2, but for the outburst of April 2, 2023.} {
\begin{tabular}{lrrrl}
\hline
DATE(UT)     & r & $\Delta$ & radius & Obs. \\
             & (au) & (au) & (km) &  \\
\hline
2024-04-03.79 & 0.85 & 1.61 & 50000.0 & T \\
2024-04-04.81 & 0.84 & 1.61 & 90000.0 & T \\
2024-04-06.80 & 0.83 & 1.61 & 172000.0 & T \\
2024-04-06.84 & 0.83 & 1.61 & 168500.0 & T \\
2024-04-07.78 & 0.82 & 1.61 & 193000.0 & T \\
2024-04-08.80 & 0.81 & 1.61 & 225000.0 & T \\
2024-04-09.44 & 0.81 & 1.61 & 249500.0 & T \\
2024-04-09.45 & 0.81 & 1.61 & 260000.0 & T \\
\hline
\end{tabular}}\label{tab:4}
\end{table}

After the first outburst of Comet 12P on July 20, 2023, a horseshoe-shaped arc structure opened to the north was observed, and it was seen gradually expanding (Figure 2). Visually, the comet nucleus appears to be located on the southern edge of this arc structure. The structure expanded while maintaining almost the same shape from July to September. Since fine dust that would form a dust tail would be blown away by solar radiation pressure, the fact that this structure remained intact for such a long time suggests that the arc structure is composed of relatively large dust particles.

\section{Outburst expansion velocities}\label{ssec:23}

By measuring the diameter of the outer edge of the dust coma, it is possible to determine the expansion velocity of the dust coma. We measured the expansion velocities for the outbursts on July 23, 2023, November 14, 2023 and April 2, 2024.
For the July 2023 outburst, the diameter was measured in the east-west direction to avoid the dust coma extending to the southeast. For the November 2023 and April 2024 outbursts, the diameter was measured in the direction perpendicular to the solar direction, and the radius was converted to an actual length.
The observations used for these measurements are summarized in Tables 2, 3, and 4. Figure 3 plots the expansion of the dust coma and the arc-shaped structure during the outburst on July 20, 2023. Through linear fitting, we were able to determine the expansion velocity and the occurrence time of the outburst. As a result, the expansion velocity of the outer dust coma was 207.5 $\pm$ 0.1 $\rm{ms}^{-1}$, and the outburst occurred at 2023-07-20.39 $\pm$ 0.05 UT. The expansion velocity of the inner arc-shaped structure was 52.0 $\pm$ 1.7 $\rm{ms}^{-1}$, and its occurrence time was 2023-07-19.54 $\pm$ 0.43 UT. These differences in expansion velocities are likely due to differences in the particle sizes of the dust and the apparent geometry.

\begin{figure}\label{fig:3}
 \begin{center}
   \includegraphics[width=8.5cm]{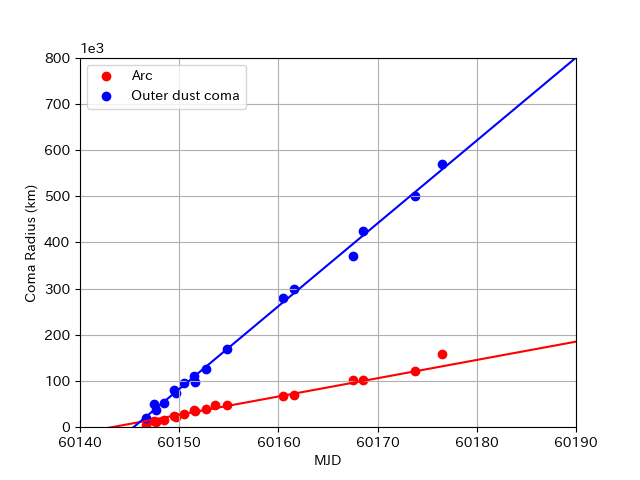} 
 \end{center}
 \caption{A plot showing the time variation in the radius of the expanding coma after the outburst in July 2023. The horizontal axis represents MJD (Modified Julian Date), where MJD=60140 corresponds to July 15, 2023. The blue plots represent the isotropically expanding outer coma, while the red plots indicate the measured radius of the arc structure.
{Alt text: Graph of coma and arc expansions of the Jul 2023 outburst}}
\end{figure}

Figure 4 plots the expansion of the dust coma during the outburst on November 14, 2023. The expansion velocity was 305.8 $\pm$ 13.7 $\rm{ms}^{-1}$, and the outburst occurred at 2023-11-14.52 $\pm$ 0.10 UT. Figure 5 shows the expansion of the dust coma during the outburst on April 2, 2024. The expansion velocity was 408.9 $\pm$ 14.3 $\rm{ms}^{-1}$, and the outburst occurred at 2024-04-02.23 $\pm$ 0.06 UT. The expansion velocity increases with decreasing heliocentric distance, likely due to higher gas and dust initial velocities as the comet nears the Sun. Table 5 summarizes the expansion velocities obtained from our observations and compares them with the results of \citep{james}.

\begin{figure}\label{fig:4}
 \begin{center}
   \includegraphics[width=8.5cm]{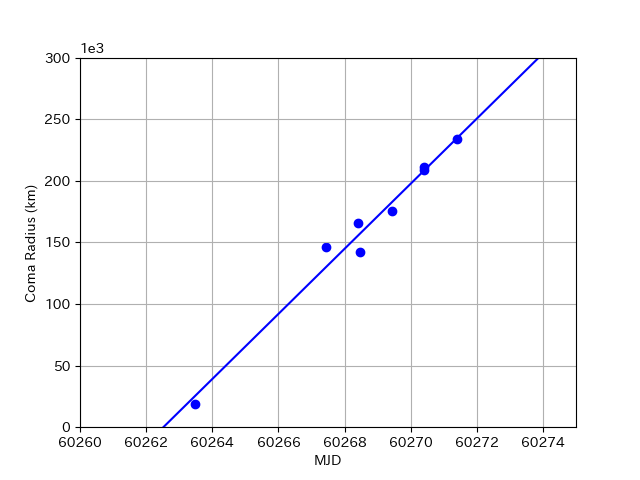} 
 \end{center}
 \caption{A plot showing the time variation in the radius of the expanding coma after the outburst in November 2023. MJD=60260 corresponds to Nov 12, 2023.
{Alt text: Graph of coma expansion of the Nov 2023 outburst}}
\end{figure}

\begin{figure}\label{fig:5}
 \begin{center}
   \includegraphics[width=8.5cm]{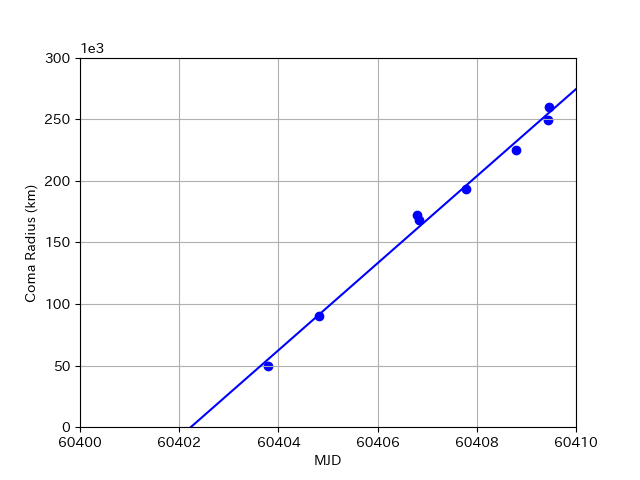} 
 \end{center}
 \caption{A plot showing the time variation in the radius of the expanding coma after the outburst in April 2024. MJD=60400 corresponds to Mar 31, 2024.
{Alt text: Graph of coma expansion of the Apr 2024 outburst}}
\end{figure}

\begin{table}
  \centering
  \caption{Expansion velocities of the outbursts. N is the numnber of observations.}
           {
  \begin{tabular}{lrlrrl}
\hline
    & velocity   & Outburst date & N & r \\
    & (ms$^{-1}$) &  (UT)        &    & (au) \\
\hline
coma & 207.5$\pm$3.1 & 2023-07-20.39$\pm$0.05 & 17 & 3.89 \\
coma$^*$ & 201 & 2023-07-20.39 & - & 3.89  \\
arc  & 52.0$\pm$1.7 & 2023-07-19.54$\pm$0.43 & 18 & 3.90 \\
coma$^*$ & 228 & 2023-10-05.16 & - & 3.07  \\
coma$^*$ & 194 & 2023-10-31.48 & - & 2.76  \\
coma & 305.8$\pm$13.7 & 2023-11-14.52$\pm$0.10 & 17 & 2.59 \\
coma$^*$ & 313 & 2023-11-14.70 & - & 2.59  \\
coma & 408.9$\pm$14.3 & 2024-04-02.23$\pm$0.06 & 17 & 0.86 \\
\hline
\end{tabular}
\begin{tabnote}
\footnotemark[$*$] Data from BAA observations (James, 2023).
\end{tabnote}
}

\label{tab:5}

\end{table}

\section{Numerical model of a dust jet}\label{sec:4}

\subsection{Constraints of latitude range of the active region}\label{ssec:41}

\begin{figure}\label{fig:6}
 \begin{center}
   \includegraphics[width=8.5cm]{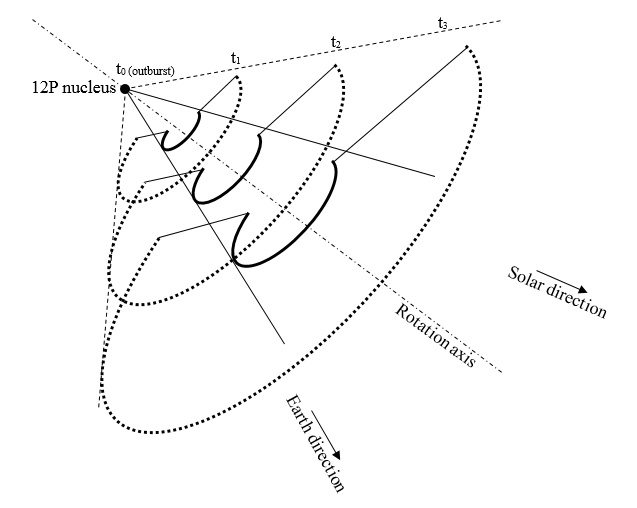} 
 \end{center}
 \caption{A schematic diagram of the comet nucleus and the arc structure. The rotation axis is angled slightly away from the line of sight from Earth to the comet nucleus. The model assumes that a dust jet ejected from the mid-to-high latitudes of the comet nucleus traces an arc due to the comet's rotation, forming a semicircular arc structure. Based on the observed images, it is considered that the arc structure is aligned with the line of sight.
{Alt text: Schematic diagram of the arc structure}}
\end{figure}

We consider the geometry of the comet nucleus and the arc structure as shown in Figure 6.
The rotation axis is oriented towards the Sun and the Earth.
Dust is ejected along with gas from an active region located at mid-latitudes, and it spreads out in a spiral due to the comet's rotation.
The spiral appears as an arc because the active period is shorter than the rotation period, revealing only part of the spiral.
The arc structure overlaps with the line of sight, causing the arc structure and the central brightness of the comet to appear coincident.
Due to this geometric configuration, the center of the arc appears to be nearly aligned with the direction of the rotation axis. Based on this assumption, we estimate the direction of the rotation axis from the arc structure. The apparent speed at which the center of the arc structure moves away from the comet is the speed projected onto the plane perpendicular to the line of sight. A higher latitude of the active region requires a faster ejection velocity, while a lower latitude results in a slower ejection velocity.
If the latitude of the active region is $\phi$, the dust ejection velocity $V_d$ can be derived from the projected expansion velocity of the arc structure $V_p$ as:

\begin{equation}
V_d = \frac{V_p}{\sin (\pi/2 - \phi)}  \tag{1}
\end{equation}

Figure 7 plots the dust ejection velocity required for each latitude of the active region based on the observed projected expansion velocity $V_p=52.0 \rm{ms}^{-1}$ during the July 20 outburst. The upper limit for the expansion velocity is set at 207.5 $\rm{ms}^{-1}$, which is the isotropic expansion velocity of the outer dust coma. From this, we estimate that the latitude of the active region during the outburst is in the range of $60 \degree  \sim 75\degree$. Assuming that the active region is within this latitude range, in the next section, we will attempt to reproduce the arc structure through dust ejection model calculations and explore the direction of the rotation axis.

The necessary dust ejection velocities for each latitude of the arc structure's projected expansion speed are summarized in Table 6. The higher the latitude of the active region, the faster the required ejection velocity. We will now explore the direction of the rotation axis that can reproduce the observed structure near this range of latitudes and ejection velocities.

\begin{figure}\label{fig:7}
 \begin{center}
   \includegraphics[width=8.5cm]{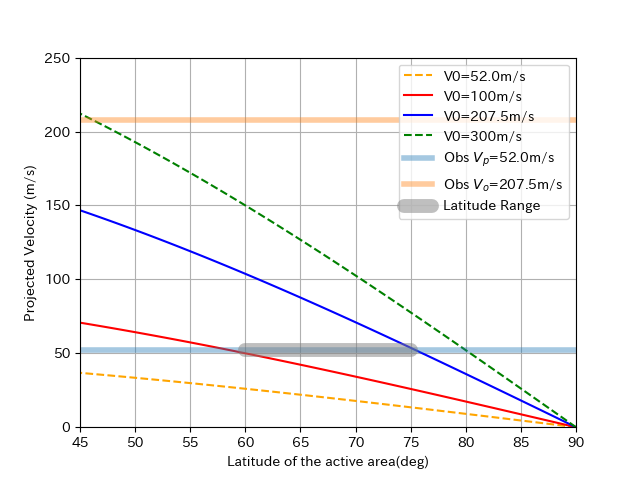} 
 \end{center}
 \caption{This figure shows the relationship between the latitude of the active region and the dust ejection velocity required to achieve the arc structure's expansion velocity of 52.0 $\rm{ms}^{-1}$ under the geometric conditions of Figure 6, with the gray-shaded area indicating the constrained region.
{Alt text: Latitude of the active spot and the dust ejection velocity}}
\end{figure}

\begin{table}
  \centering
  \caption{The latitude $\phi$ of the outburst and the dust ejection velocity $V_d$ required for the projected expansion speed of the arc structure.}{
  \begin{tabular}{rr}
    \hline
    Latitude \(\phi\) & \(V_d\) \\
    ($\degree$) &($\rm{ms}^{-1}$) \\
    \hline
60 & 104 \\
65 & 123 \\
70 & 152 \\
75 & 201 \\
80 & 300 \\
    \hline
  \end{tabular}}\label{tab:6}
\end{table}

\subsection{Dust ejection velocity}\label{sec:42}

The dust ejection velocity is determined by its interaction with the gas being emitted. At the heliocentric distance of 4 au, which triggered the outburst in July 2023, $\rm{H}_2\rm{O}$ was not yet active, so $\rm{CO}$ or $\rm{CO}_2$ is considered to be the main gas component. Here, we consider $\rm{CO}$ as the driving force of the outburst and use a modified model of \citet{probstein} \citep{wallis} to determine the terminal velocity of the dust $V_d$. The ratio of dust to gas ejection velocities $V_d/V_g$ is expressed by the following equation:

\begin{equation}
V_d/V_g = [0.9 + 0.4\mu^{1/2} + 1.5(a/\Lambda)^{1/2}]^{-1} \tag{2}
\end{equation}

\begin{equation}
\Lambda = 2 m_w Q / \rho_d R V_g \tag{3}
\end{equation}

\begin{eqnarray*}
    m_w :& \text{molecular weight, for } \rm{CO} = 4.65\times10^{-26} \rm{kg}\\
    \mu :& \text{dust mass ratio to total outflow} \\
    Q   :& \text{outflow rate (per second and steradian)}\\
    \rho_d :& \text{dust density},  800 \text{kg m}^{-3} \text{\citep{wallis}}\\
    V_g :& \text{gas expansion velocity} \\
    R  :& \text{radius of the nucleus, } 17 \rm{km} \text{\citep{ye}}\\
\end{eqnarray*}

For the dust-to-gas ratio $\mu$, we adopt $\mu = Q_d/(Q_d + Q_g) \sim 0.5$, based on $Q_d/Q_g \sim 0.84$ as determined for Comet 67P/Churyumov-Gerasimenko \citep{marschall}. Since there are no observations of the $\rm{CO}$ production rate for Comet 12P, we estimate the $\rm{CO}$ production rate using the relationship between the $\rm{CO}$ production rate and the visual R-band magnitude $m_R$ of Comet 29P/Schwassmann-Wachmann 1, 
as observed by \citet{bockelee}. They showed that there is a relationship between the $\rm{CO}$ production rate of Comet 29P and $m_R$:

\begin{equation}
\log_{10}(Q(\rm{CO})_{total}) = (29.29 \pm 0.04) - (0.062 \pm 0.004)m_R(1,r_h,0) \tag{4}
\end{equation}

\begin{equation}
m_R(1, r_h, 0) = m_R - 5 \log\Delta  \tag{5}\\
\end{equation}

Using this relationship, and substituting the values $m_R=11.0$, $r_h = 3.89 \rm{au}$, $\Delta=3.58 \rm{au}$ for the outburst of Comet 12P on Jul 20, 2023, the total production rate $Q(\rm{CO})_{total}$ is estimated to be:

\begin{equation}
Q(\rm{CO})_{total} \sim 6.0\times10^{28} s^{-1} \tag{6}
\end{equation}

From observations of Comet 29P, we estimate that the ratio of jet components to the total $Q(\rm{CO})_{total}$ is about $\sim 0.5$ \citet{bockelee}, so the $\rm{CO}$ production rate of the outburst component can be roughly estimated as follows:

\begin{equation}
Q(\rm{CO})_{jet} \sim Q(\rm{CO})_{total}\times 0.5 \sim 3.0\times10^{28} s^{-1} \tag{7}
\end{equation}

The gas ejection velocity $V_g$ is based on the blue-shift velocity of 480 $\rm{ms}^{-1}$ observed during a burst of Comet 29P at a heliocentric distance of 6.17 au (Crovisier et al., 1995), and we apply a heliocentric distance dependence proportional to $r^{-0.5}$.

\begin{equation}
  V_g \sim 1200 r^{-0.5} \rm{ms}^{-1} \tag{8}
\end{equation}

At the time of the outburst on July 20, 2023 ($r=3.89\rm{au}$), $V_g=608\rm{ms}^{-1}$. The $\rm{CO}$ production rate per unit solid angle $Q$ is determined by the ratio $f\sim 0.1$ for the active region of the outburst being hemispherical, and substituting $Q = Q(\rm{CO})_{jet}/f$, we obtain $\Lambda = 3.4 \micron$. If the dust radius $a < \Lambda$, it is affected by the gas velocity.

\begin{figure}\label{fig:8}
 \begin{center}
   \includegraphics[width=8.5cm]{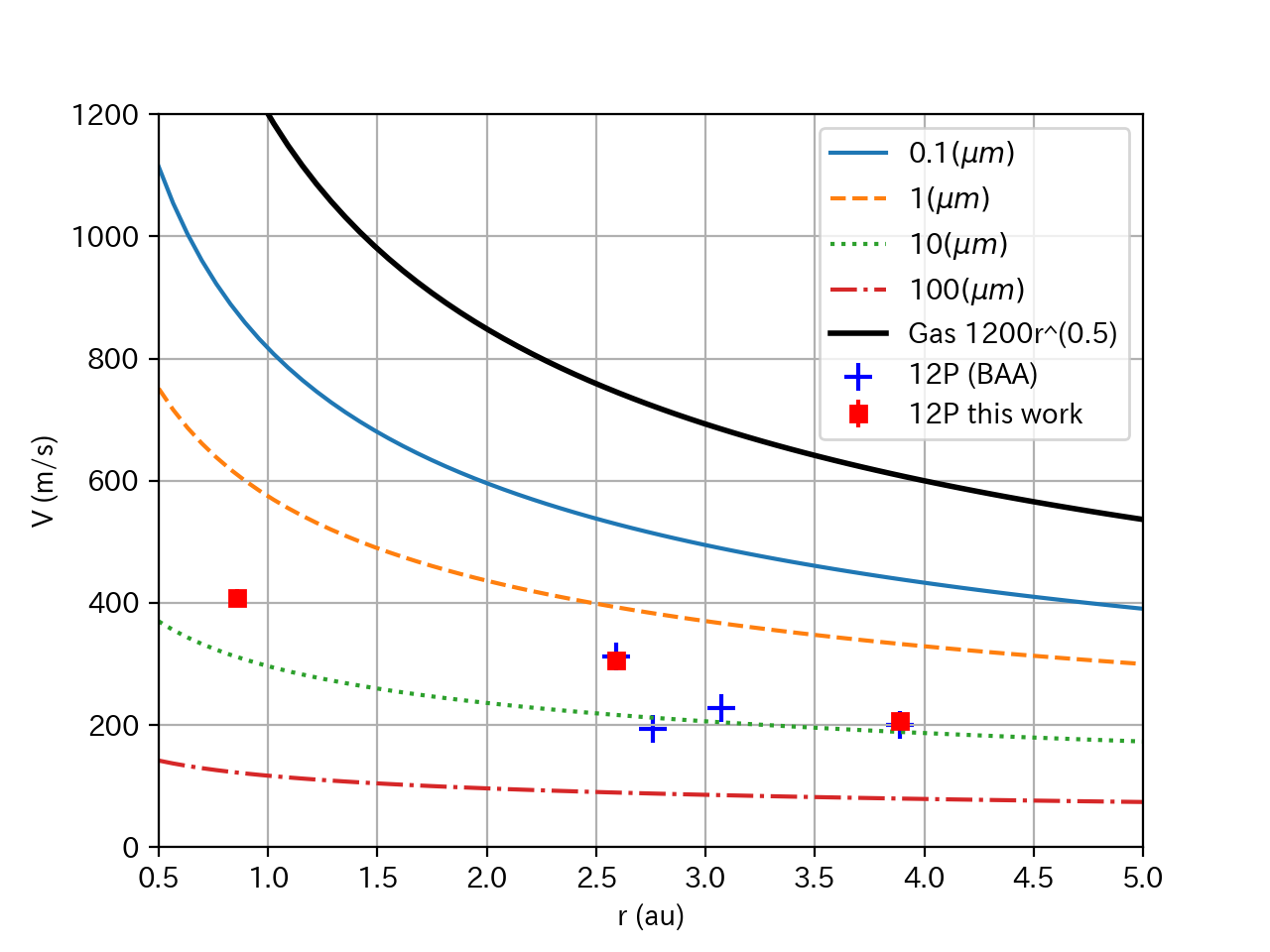} 
 \end{center}
 \caption{The heliocentric distance dependence of ejection velocity for each dust size and the observed dust coma expansion velocity. The size of the dust during the outburst is estimated to be around $10 \micron$. The BAA observations are from \citet{james}.
{Alt text: Heliocentric distance and dust ejection velocity}}
\end{figure}

Figure 8 shows the heliocentric distance dependence of ejection velocities for each dust size. Since the observed dust coma expansion velocity is around 200 $\rm{ms}^{-1}$ at the July outburst, the size of the emitted dust particles is estimated to be between 1 and 10 $\mu$m.

\begin{table}
  \caption{Parameters used in the model calculations}{
  \begin{tabular}{lrl}
    \hline
    Parameter & Value & Refs \\
    \hline
radius of the nucleus & 17\(\pm6\) km & \citet{ye} \\
onset of outburst & Jul 20.39, 2023 & \\
\(\beta\) & 0.1 & \\
Rotation period & 2.38 day & \citet{knight} \\
half cone angle & 10\degree & \\
Orbital elements & & MPEC 2023-R20 \\
    \hline
  \end{tabular}}\label{tab:7}
\end{table}

\subsection{Model calculations of the arc structure}\label{sec:4.3}

The dust coma consists of an isotropically expanding component and an arc structure component. Here, we developed a numerical model to obtain information on the direction of the comet nucleus's rotation axis and the latitude of the active region from the arc structure caused by the localized outburst.

To reproduce the observed arc structure, we developed a model that tracks the motion of dust particles ejected from a specific location on the comet nucleus through numerical integration. Dust is accelerated by the flow of gas sublimating from a specific location (active region) on the comet, and moves away from the comet nucleus at terminal velocities that vary according to particle size. We numerically integrate the motion of individual dust particles, taking into account solar radiation pressure, and compare their distribution with the observed structure. The initial velocity of the dust is the radial velocity at which it is ejected from the surface, combined with the comet's orbital motion and the rotational motion of the comet nucleus (\cite{probstein}).

\begin{equation}
\frac{d\boldsymbol{v}}{dt} =  \frac{GM_\odot}{r^2}\frac{\boldsymbol{r}}{r} (1 - \beta) \tag{8}
\end{equation}
\begin{equation}
\beta = \frac{F_{rad}}{F_{grav}} = \frac{C Q_{pr}}{2 \rho_d a} \tag{10}
\end{equation}

Here, $\mathbf{v}$ is the velocity of the dust, $\mathbf{r}$ is the position of the dust, $r=|\mathbf{r}|$, $F_{rad}$ is the solar radiation pressure, and $F_{grav}$ is the solar gravitational force. $\beta$ is the ratio of these forces, and the larger the $\beta$, the more the dust particles are affected by solar radiation pressure, causing their orbital motion to be pushed outward. $a$ is the diameter of the dust, $\rho_d$ is the dust density $\sim 800 (\rm{kg}/m^3)$, $G$ is the gravitational constant, $M_\odot$ is the mass of the Sun, $c$ is the speed of light, and $C$ is a constant independent of the heliocentric distance, with $C=1.19\times10^{-3} (\rm{kg}/m^2)$. $Q_{pr}$ is the scattering efficiency, and if $a \geq 1{\mu m}$, it can be approximated as $\sim 1$. From the analysis of dust tails, $\beta$ is found to range from $10^{-5} \sim 1.0$ \citep{fulle}. Small dust particles with $\beta > 1$ are blown away in the antisolar direction by solar radiation pressure and do not remain near the nucleus. Dust that stays near the nucleus for an extended period, like the arc structure of Comet 12P, is thought to be composed of larger particles with $\beta \sim 0.1$ or less.

\begin{figure*}\label{fig:9}
 \begin{center}
   \includegraphics[width=18cm]{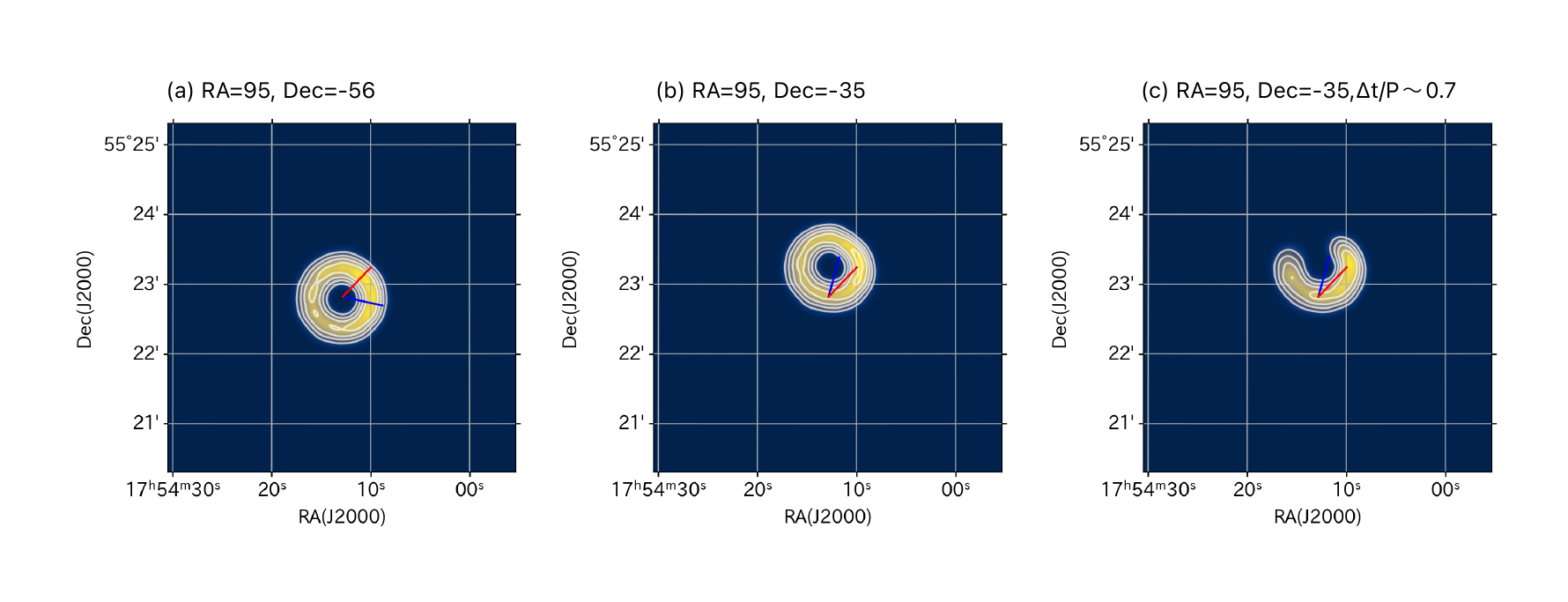} 
 \end{center}
 \caption{Differences in the appearance of the arc structure formed by the dust jet ejected from a latitude of $70\degree$, depending on the direction of the rotation axis. The blue and red lines represent the rotation axis and the solar direction, respectively.
(a) When the line of sight (RA=$95\degree$, Dec=$-56\degree$) coincides with the rotation axis, and the outburst active period is the same as the rotation period. (b) When the declination of the rotation axis is set to $-35\degree$. The arc structure overlaps with the comet nucleus. (c) When the ratio of the outburst active period to the rotation period is set to 0.7.
{Alt text: Appearance of dust coma strurcture for different rotation axes}}
\end{figure*}

While the active region is illuminated by sunlight, dust is ejected at regular intervals in random directions within a cone specified by a half-angle, with a velocity $V_d$.
The variation in outburst intensity with local time is assumed to be
proportional to the cosine of the zenith angle of the active region.
However, as discussed later, the estimated orientation of the spin
axis and the latitude of the active region place it in the summer
hemisphere, where it is continuously illuminated by the Sun, thereby
reducing the influence of local time effects.  The duration of
activity was estimated as a fraction of the rotation period, chosen to
reproduce the extent of the observed arc structure.
The initial velocity is determined by adding the rotational angular velocity and the orbital motion of the comet. Using the 5th-order Runge-Kutta method, numerical integration is performed to calculate the dust distribution after a certain period, and the results are compared with the observed structure of the dust coma. The rotation period was adopted as a parameter, based on the period of $57\pm 1$ hours (=2.38 days) inferred by \citet{knight} from the spiral structure observed between January 30 and February 16, 2024. The 10-degree half-angle of the jet cone was chosen to prevent the arc structure from becoming indistinct, and falls within the range of 10 to 35 degrees observed for jets in Comet 67P/Churyumov-Gerasimenko during the ROSETTA mission \citep{lai}.


We estimated that the active region of the outburst is located at a latitude in the range of 60$\degree$ to 75$\degree$. From the positional relationship between Earth and the comet during the outburst on July 20, 2023, the right ascension and declination of Earth as seen from the comet (line of sight) are RA = $95\degree$, Dec = $-56\degree$ (J2000.0). Assuming that the rotation axis is near this direction, we examine the relationship between the dust distribution spiraling out due to rotation and the direction of the rotation axis, fixing the dust ejection velocity and $\beta$.

For dust with an active region latitude of $70\degree$, a dust ejection velocity of
$V_d=152\ \rm{ms}^{-1}$, and $\beta=0.1$, we calculate the dust distribution on August 5, 2023, by varying the rotation axis and examine the conditions that reproduce the arc structure. The positions of the dust particles are projected onto a 5-arcminute field of view on the celestial sphere, and the distribution is represented as the number of dust particles on a $256\times256$ grid centered on the comet nucleus.

When the rotation period is $P$ and the active duration of the burst is $\Delta t$, the ratio of the two determines the the resulting structure. Specifically, when $\Delta t/P << 1$, a linear jet structure forms. As the ratio increases to $\Delta t/P < 1$, the structure takes the shape of an arc. When $\Delta t/P \sim 1$, it forms a nearly complete ring, and for $\Delta t /P >> 1$, a multi-spiral structure emerges.

Figure 9(a) shows the model calculation result when the line of sight coincides with the rotation axis, with the rotation period and the outburst active period set to the same value (2.38 days), resulting in a ring-shaped distribution centered on the comet nucleus. Since the observed arc structure overlaps with the comet itself, Figure 9(b) shows the result when the declination of the rotation axis is changed to $-35\degree$ so that the southern part of the ring overlaps with the comet nucleus. In the observations, the northern part of the ring is missing, so the outburst active period must be shorter than the rotation period. If the rotation period of Comet 12P is 2.38 days, as determined by Knight et al. (2024), the outburst active period would be about 1.7 days. Figure 9(c) shows the dust distribution when the active period is set to $\Delta t=1.7$ days. It can be seen that the observed arc structure is well reproduced when the ratio of the outburst active period to the rotation period is $\Delta t/P \sim 0.7$. This ratio is independent of the rotation period; if the rotation period is shorter, the active period also becomes shorter.

\begin{figure}\label{fig:10}
 \begin{center}
   \includegraphics[width=8.5cm]{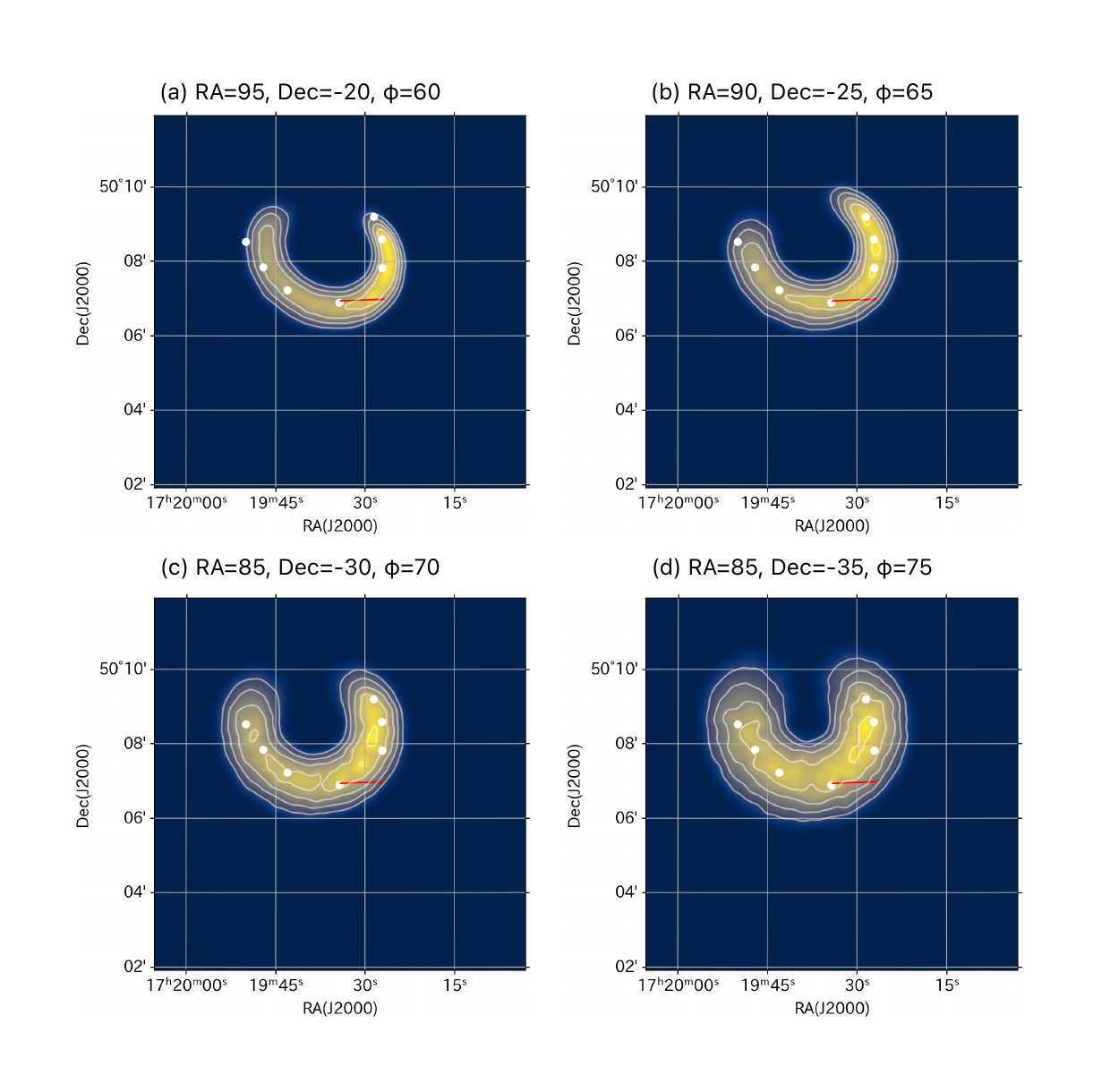} 
 \end{center}
 \caption{Model calculation results for September 13, 2023, with varying latitudes $\phi$ of the outburst active region. The white dots plot the ridge line of the arc structure based on the contours of the observed images. The red line represents the solar direction.
If the latitude of the active region is in the range of $60\degree \sim 75\degree$, the observed arc structure can be reproduced by setting the direction of the rotation axis within the range of $85\degree < \rm{RA} < 95\degree$ and $-35\degree < \rm{Dec} < -20\degree$.
{Alt text: Model calculation results for various rotation axes}}
\end{figure}

Figure 10 compares the modeled dust distribution for September 13, 2023, with the measured points of the observed arc structure when the latitude of the active region is set to 60°, 65°, 70°, and 75°s. The dust ejection velocities for each latitude are taken from the values obtained in Table 6 based on geometric constraints. The calculated field of view is 10 arcminutes on the celestial sphere. The outburst active period is set to 1.7 days from the onset. The positions of the brightness peaks of the arc structure on the same day, as measured from observations, are overlaid as white dots on the model calculation results. This is the result of calculating the direction of the rotation axis in $5\degree$ increments for each active region latitude that best matches the observed white points.

The dust ejection velocity $V_d$ and the right ascension and declination of the rotation axis for each active region latitude are summarized in Table 8. The observed arc structure can be reproduced over a period of two months if the direction of the rotation axis is within the range RA=$85\degree \sim 95\degree$ and $-20\degree < \rm{Dec} < -35\degree$.

It is important to estimate the size of the dust that constitutes the observed arc structure. Figure 11 shows the results of varying the value of $\beta$ for the dust distribution on September 13, 2023, assuming $\phi=70\degree$, $\rm{RA}=85\degree$, $\rm{Dec}=-30\degree$, and $V_d=152\ \rm{ms}^{-1}$. When $\beta > 0.1$, the dust is seen being pushed in the antisolar direction by solar radiation pressure. The dust that forms the arc structure, which maintains its shape for nearly two months, has a small $\beta$, indicating that it consists of larger dust particles.

\begin{figure*}\label{fig:11}
 \begin{center}
   \includegraphics[width=18cm]{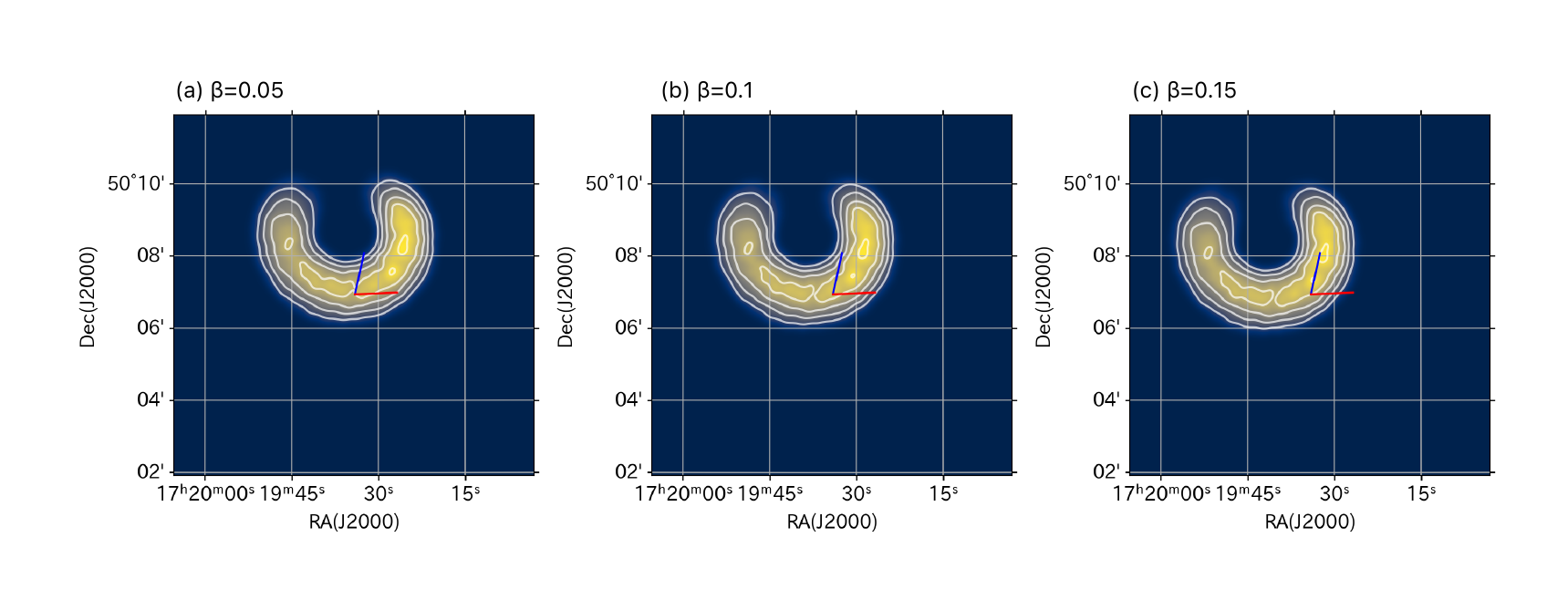} 
 \end{center}
 \caption{Results calculated for $\beta=0.05, 0.1, 0.15$, replicating the observations from September 13, 2023. It can be seen that for $\beta > 0.1$, the arc structure is pushed in the antisolar direction (eastward) by radiation pressure. The blue and red lines represent the rotation axis and the solar direction, respectively.
{Alt text: Model calculation results for various $\beta$}}
\end{figure*}

Figure 12 shows the position angles of the dust ejected in the direction of the rotation axis within the candidate range, compared with the position angles of the center of the arc structure measured from images. If the direction of the rotation axis is at the center of the arc structure, the position angle around mid-August 2023 (MJD=60170$\sim$60180) cannot be reproduced for RA < 90. From this, it is considered that the rotation axis is located near RA $\sim 85\degree$ and $-30\degree < \rm{Dec} < -35\degree$.

\begin{figure}\label{fig:12}
 \begin{center}
   \includegraphics[width=8.5cm]{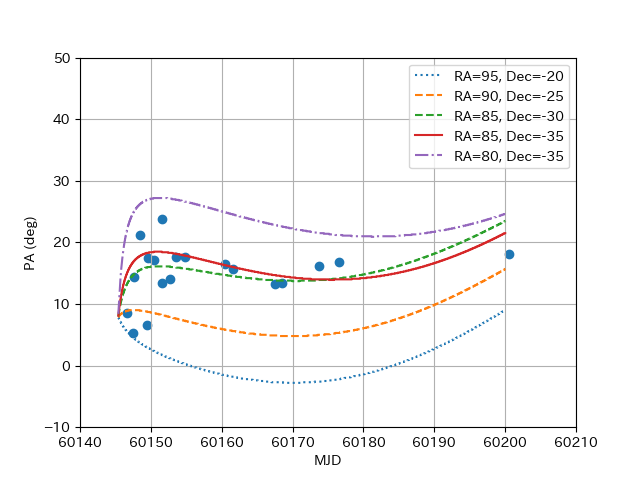} 
 \end{center}
 \caption{    The position angle of the center direction of the arc structure
    measured from images, and the time variation of the position angle of
    each rotation axis direction from the model calculations. When $\rm{RA} > 90\degree$
    or $\rm{RA} < 80\degree$, the model does not match the observations around
    mid-August (MJD$\sim$60170). It can be seen that $\rm{RA}\sim 85\degree$ matches the
    observations well.
    {Alt text: Observed and calculated position angles of the arc structure}}
\end{figure}

\begin{table}
\centering
  \caption{The dust ejection velocity for each latitude of the active region and the right ascension and declination of the rotation axis.}{
  \begin{tabular}{rrrr}
    \hline
Latitude & $V_d$ & RA & Dec \\
($\degree$) & ($\rm{ms}^{-1}$) & ($\degree$) & ($\degree$)\\
    \hline
60 & 104 & 95 & -20 \\
65 & 123 & 90 & -25 \\
70 & 152 & 85 & -30 \\
75 & 201 & 85 & -35 \\
    \hline
  \end{tabular}}\label{tab:8}
\end{table}


So far, model calculations have been conducted for dust particles of the same size. It is known that the dust ejected from the comet nucleus follows an inverse power-law size distribution. According to \citet{fulle}, the size distribution index $\alpha$ for several comets, based on the analysis of dust tails, was in the range $-4.0 > \alpha > -3.0$. Here, as an average value, the dust size distribution $N(a)$ is given by:

\begin{equation}
  N(a) \propto a^{-3.5} \tag{11}
\end{equation}

Model calculations were performed where dust was ejected from the comet nucleus in quantities following this size distribution. The initial values for the dust ejection velocity were set according to the size-dependent values given by equation (11).

\begin{figure}\label{fig:13}
 \begin{center}
   \includegraphics[width=8.5cm]{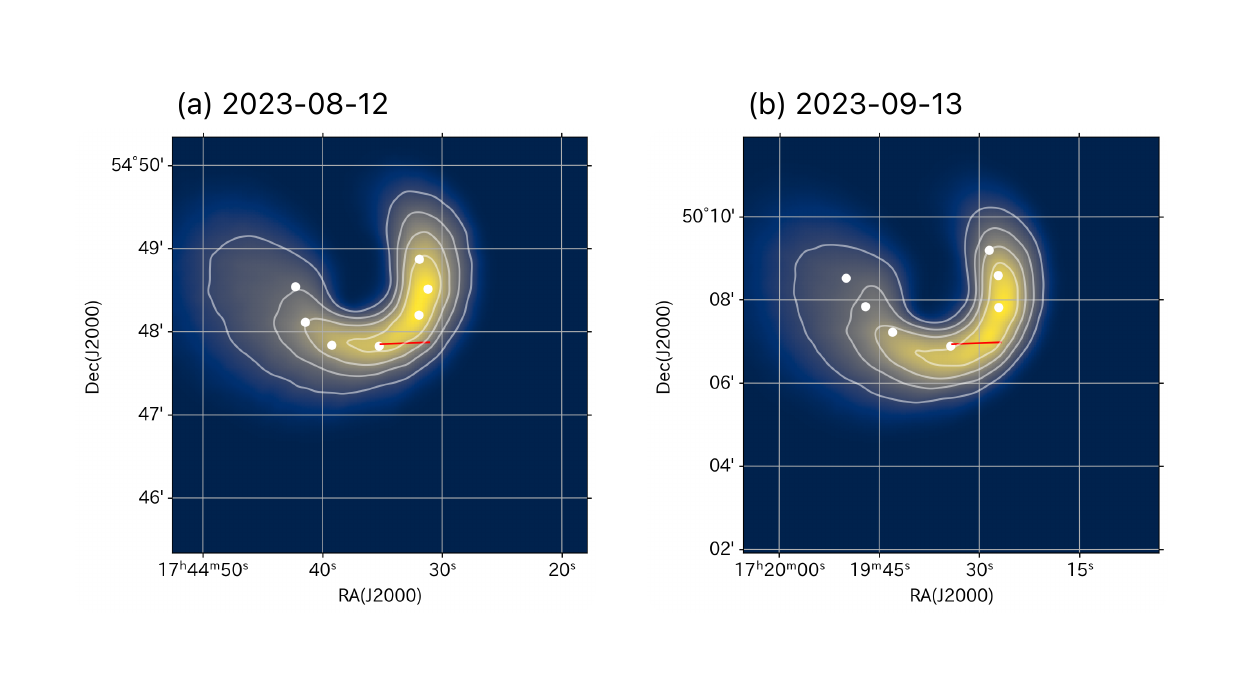} 
 \end{center}
 \caption{(a) is the model calculation result for the dust size distribution on August 12, 2023, with $\phi=70\degree$, $\rm{RA}=85\degree$, $\rm{Dec}=-35\degree$, and dust radius calculated for $1 < a < 20 \micron$. (b) is the model calculation result for the dust distribution on September 13, 2023, with $\phi=70\degree$, $\rm{RA}=85\degree$, $\rm{Dec}=-35\degree$, and dust radius calculated for $3 < a < 30 \micron$.
{Alt text: Model calculation results including dust size distribution}}
\end{figure}

Model calculations were conducted in which dust was ejected from the comet nucleus in quantities following the size distribution. The initial values for the dust ejection velocity were set according to size-dependent values based on equation (2). The ratio of solar radiation pressure to gravity $\beta$ for each dust size was calculated using equation (9), and the sum of the values, weighted by the scattering cross-section of the dust, was used to create a map represented by pixel values.

Figure 13 shows a comparison between the model calculation results, assuming a rotation axis direction of RA=$85\degree$, Dec=$-35\degree$, and an active region latitude of 70°, and the measurements of the arc structure. The model successfully reproduces the behavior of smaller dust particles being pushed in the antisolar direction, resulting in a structure that more closely matches the observed one.

For the calculations on August 12, 2023, the dust size range was set to $1 < a < 20\micron$, and for the calculations on September 13, 2023, it was set to $3 < a < 30\micron$. By September 13, two months after the outburst, much of the smaller dust had been blown away by solar radiation pressure, leaving the larger dust particles as the main contributors.

\section{Discussion}\label{sec:5}

\begin{figure}[t]\label{fig:14}
 \begin{center}
   \includegraphics[width=8.5cm]{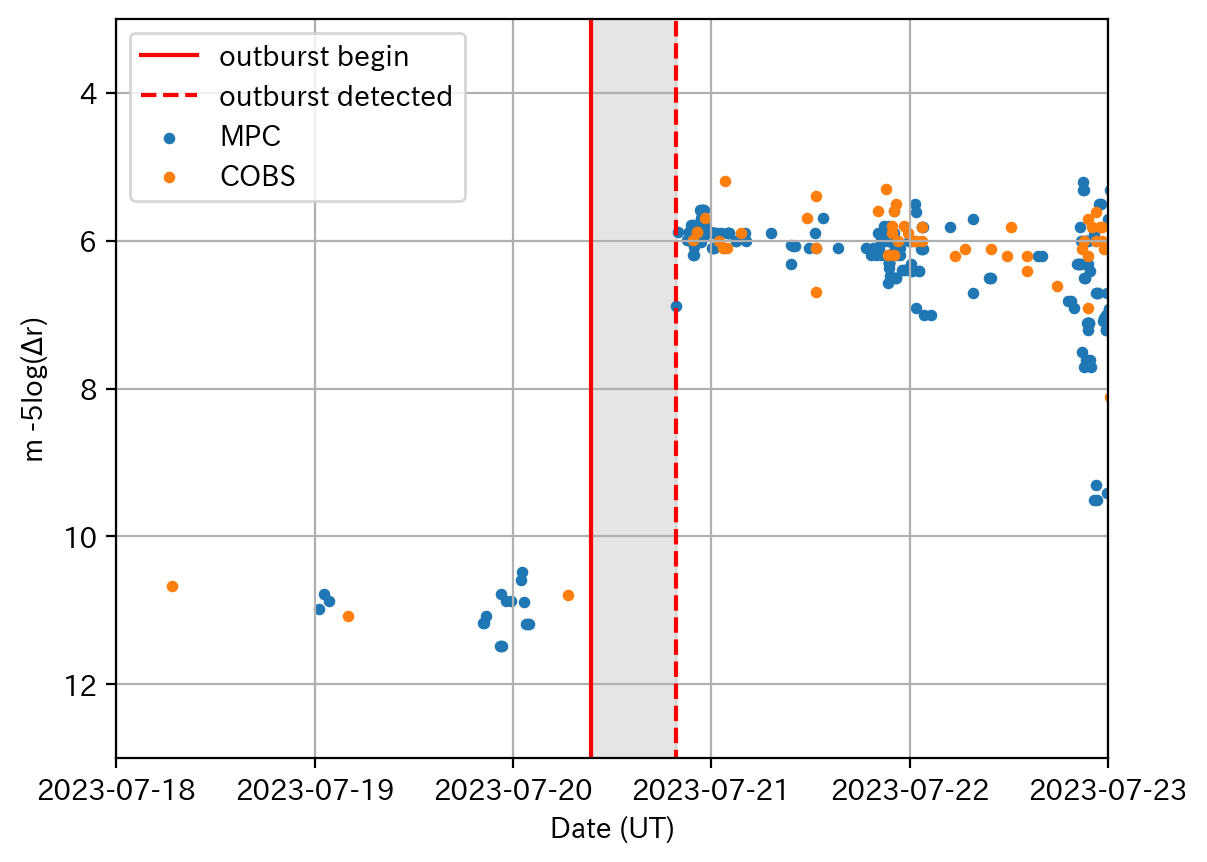} 
 \end{center}
 \caption{The brightness variation of Comet 12P before and after the outburst on July 20, 2023 (from MPC and COBS). The time from the occurrence of the outburst to its detection was 0.43 days. This time represents the upper limit of the outburst's active period.
{Alt text: Brightnes variation around the July 2023 outburst}}
\end{figure}

In modeling the arc structure created by the outburst, we used the rotation period of 2.38 days (57 hours) determined by \citet{knight}. This rotation period was derived from the reproduction of the spiral structure observed from January to February 2024. On the other hand, model calculations showed that for the arc structure to form due to rotation, the ratio of the outburst active period to the rotation period needs to be around 0.7. If the rotation period is 57 hours, the required outburst active period would be approximately 40 hours (1.7 days).

Figure 14 shows the brightness observations before and after the outburst on July 20, 2023, along with the estimated outburst occurrence time (July 20.39, 2023) based on the dust coma expansion. The first observed brightening after the estimated outburst occurrence time was 12.6 magnitude at Jul 20.82, 2023 (MPC), and since no further brightening was observed after that, it is likely that the outburst had already ended, and the ejection of dust from the nucleus surface had ceased.

From the above, the period of 0.43 days between the occurrence of the outburst (July 20.39, 2013) and the confirmation of the brightening (July 20.82, 2023) can be considered the upper limit of the outburst's active period. Assuming a rotation period of 2.38 days, 0.43 days is insufficient to create the arc structure. For the arc structure to form within an active period of 0.43 days, the rotation period of Comet 12P would need to be around 14 to 15 hours. If the rotation period is assumed to be 14.3 hours, it would be one-third of 57 hours, making it possible to also reproduce the spiral pattern.

Model calculations with fixed dust sizes showed that the dust constituting the arc structure formed by the outburst has $\beta \lesssim 0.1$. Converting this to dust radius using equation (9), the size is estimated to be $a \gtsim 15\micron$. On the other hand, the dust size derived from the dust ejection velocity model (Figure 8) is around $\sim 10\micron$, which is the same order of the estimate based on the value of $\beta$. To maintain the shape of the arc structure against solar radiation pressure over two months after the outburst, it is likely that two factors played a role: the direction of the rotation axis and the size of the dust particles.

In our model calculations, we set the cone angle of the dust jet ejected from the active region to $20\degree$ (half cone angle = $10\degree$). Increasing this angle makes the arc structure less distinct, and to maintain the observed arc structure over two months, it needs to be around 10°. This angle falls within the range of $10\degree$ to $35\degree$ for the jet cone angle observed by the OSIRIS instrument on the ROSETTA mission for Comet 67P/Churyumov-Gerasimenko (\cite{lai}). From these findings, it is considered that the dust jet that formed the arc structure was ejected from a relatively narrow active region. Additionally, to form a single arc structure, the outburst must have originated from a single active region.

\section{Summary}\label{sec:6}

The paper focuses on the observation and modeling of multiple outbursts of Comet 12P/Pons-Brooks, particularly the significant outburst in July 2023. The authors tracked the expansion of the dust coma and an arc-shaped structure over several months, using continuous imaging observations. The study determined the dust coma expansion velocities for outbursts occurring on July 20, 2023, November 14, 2023, and April 2, 2024. The expansion velocities were found to be 207.5 $\rm{ms}^{-1}$, 305.8 $\rm{ms}^{-1}$, and 408.9 $\rm{ms}^{-1}$, respectively.

A key focus was the semicircular arc structure observed during the July 2023 outburst, which expanded at a slower velocity of 52.0 $\rm{ms}^{-1}$ and maintained its shape for about two months. The paper proposes that this structure was formed by dust ejected from a specific active region on the comet's nucleus, with the rotation axis nearly aligned with the line of sight.

The modeling showed that the arc structure could be accurately reproduced by assuming a rotation axis direction of RA = $85\degree$, Dec = $-30 \sim -35\degree$, with the active region located at a latitude of $70\degree$. The dust responsible for the arc structure is estimated to be larger, with a radiation pressure-to-gravity ratio ($\beta$) of approximately 0.1, indicating resistance to solar radiation pressure.

Further, the study discusses the implications of the rotation period and active duration of the outburst. The findings suggest that a shorter rotation period (~14-15 hours) is necessary to form the arc structure within the observed active period. The paper also emphasizes the role of dust size distribution in shaping the observed structures, with larger dust particles being crucial in maintaining the arc structure against solar radiation pressure.


\begin{ack}
  We acknowledge with thanks the comet observations from the COBS
  Comet Observation Database contributed by observers worldwide and
  used in this research.
  This research has made use of data and/or services provided by the
  International Astronomical Union's Minor Planet Center.
  The author would like to thank N. Ukita, who carefully read the manuscript and made valuable comments.
The authors thank the anonymous referee for their constructive comments and suggestions, which helped improve the quality of this paper.
\end{ack}



\end{document}